\documentstyle[sprocl,twoside,epsfig]{article}

\def\Journal#1#2#3#4{{#1} {\bf #2}, #3 (#4)}

\def\PLB{{\em Phys. Lett.}  B}

\def\beq{\begin{equation}}
\def\eeq{\end{equation}}
\def\bea{\begin{eqnarray}}
\def\eea{\end{eqnarray}}
\def\bq{\begin{quote}}
\def\eq{\end{quote}}

\def\rar{\rightarrow}

\def\ba{\begin{array}}
\def\ea{\end{array}}

\newcommand{\aaa}{a}
\newcommand{\dd}{{\mathrm{d}}}
\newcommand{\cZ}{{\cal Z}}

\begin{document}
\sloppy

\title{Hidden Mass Hierarchy in QCD}

\author{ V.I. ZAKHAROV }

\address{
Max-Planck Institut f\"ur Physik, F\"ohringer Ring 6, 80805 M\"unchen, Germany}

\maketitle\abstracts{We discuss implications of the recent measurements
of the non-Abelian action density associated with the monopoles condensed 
in the confining phase of gluodynamics. The radius of the monopole 
determined in terms of the action was found to be small
numerically.
As far as the condensation of the monopoles is described in terms
of a scalar field, a fine tuning is then implied.
In other words,
a hierarchy exists between the self energy of the monopole and the temperature of the
confinement-deconfinement phase transition.
The ratio of the two scales is no less than a factor of 10. 
Moreover, we argue that the hierarchy scale can well
eventually extend to a few hundred GeV on the ultraviolet side.
The corresponding phenomenology is discussed, mostly
within the polymer picture of the monopole condensation.
}

\vspace{0.4cm}


\section{Introduction}

The monopole condensation is one of the most favored mechanisms 
\cite{classic} of
the confinement, for review see, e.g., \cite{review1}.
In the field theoretical language, one usually thinks in terms
of a Higgs-type model:
\beq\label{effective}
S_{eff}~=~\int d^4x\big(|D_{\mu}\phi|^2+{1\over 4}F_{\mu\nu}^2+~V(|\phi|^2)\big)
\eeq
where $\phi$ is a scalar field with a non-zero magnetic charge,
$F_{\mu\nu}$ is the field strength tensor constructed on the  dual-gluon
field $B_{\mu}$, $D_{\mu}$ is the covariant derivative 
with respect to the dual gluon. Finally,
$V(|\phi|^2)$ is the potential energy ensuring that
$<\phi>\neq 0$ in the vacuum. Relation of the ``effective''
fields $\phi,B_{\mu}$ to the fundamental QCD fields is one of the basic
problems of the approach considered but here we would
simply refer the reader to Ref. \cite{main} for further discussion of
this problem.
At this moment, it suffices to say that the ``dual-superconductor''
mechanism of confinement assumes formation of an Abrikosov-type tube between
the heavy quarks introduced into the vacuum via the Wilson loop
while the tube itself is a classical solution of the equations of motion
corresponding to the effective Lagrangian (\ref{effective}). 

By introducing scalar fields, one opens a door to the standard questions on
the consistency, on the quantum level, of a $\lambda\phi^4$ theory.
Here, we mean primarily the problem of
the quadratic divergence in the scalar mass. At first sight, these
problems are not serious in our case since 
(\ref{effective}) apparently represents
an effective theory presumably valid for a limited range of mass scales.

However, if we ask ourselves, what are the actual limitations
on the use of the effective theory (\ref{effective}) we should 
admit that there is no way at the moment to
answer this question on pure theoretical grounds and we should
turn instead to the experimental data, that is lattice measurements. 
This lack of understanding concerns first of all the
nature of the non-perturbative field configurations that 
are defined as monopoles.
First, it is not clear apriori which $U(1)$ subgroup of
the $SU(2)$ \footnote{for simplicity we will confine ourselves 
to the case of  $SU(2)$ as the color group.} is to be picked up
for the classification of the monopoles. Even if we make this
choice on pure pragmatic basis and concentrate on the
most successful scheme of the monopoles  in the maximal
Abelian projection  \cite{review1} we still get very
little understanding of the field configurations underlying 
the objects defined as monopoles in this projection,
for discussion see, e.g., \cite{alive}.
In particular, nothing can be said on the size of the monopole
which presumably limits application of (\ref{effective}) on the
ultraviolet side.

Direct measurements of the monopole size were reported recently \cite{anatomy}
and brought an unexpectedly small value of the monopole radius:
\beq\label{size}
R_{mon}~\approx~0.06~ \mathrm{fm},
\eeq
where the monopole radius is defined here in terms of the
full non-Abelian action associated with the monopole
and not in terms of the projected action.
If we compare the radius (\ref{size}) with the temperature of
the confinement-deconfinement transition:
\beq\label{temperature}
 T_{deconf}~\approx~300~ \mathrm{MeV}
\eeq
then we would come to the conclusion that there are different mass scales coexisting within
the effective scalar-field theory (\ref{effective}). And the question,
how this mass hierarchy is maintained is becoming legitimate.

Although comparison of (\ref{size}) and (\ref{temperature})
is instructive by itself, we will argue that
the actual hierarchy mass scale can be much higher on the ultraviolet side.
Namely, we will emphasize later
that even at the size (\ref{size}) the monopoles are very ``hot'', i.e. have 
action comparable to the action of the zero-point fluctuations. 
For physical interpretation, it is natural to understand by the radius such distances
where the non-perturbative fields die away on the scale of
pure perturbative fluctuations. And this radius is to be
considerably smaller than (\ref{size}).

Also, estimate (\ref{size}) means that the asymptotic freedom is not 
yet reached at quite small distances and the question arises as to
how reconcile this observation with such phenomena as
the precocious scaling.

We cannot claim at all understanding answers to these
questions but feel that it is important to start discussing them.
Our approach is mostly phenomenological and we are trying to formulate
which measurements could help to find answers to the puzzles
outlined above. The theoretical framework which we are using is mainly the 
polymer approach to the scalar field theory, see, e.g., 
Refs.~\cite{symanzik,stone,caracciolo}.

\section{Monopole condensation: overview of the theory}

\subsection{Compact $U(1)$}

The show case of the monopole condensation is the compact $U(1)$  \cite{polyakov}.
The crucial role of the compactness is to ensure that
the Dirac string does not cost energy (for a review see, e.g.,
\cite{alive}).
The monopole self energy reduces then to the energy 
associated with the radial magnetic field
${\bf B}$.
The self energy is readily seen to diverge linearly in the ultraviolet:
\beq
\label{linear}
M_{mon}(a)~=~{1\over 8\pi}\int{\bf B}^2d^3r~\sim~{c\over 8e^2}{1\over
\aaa}\,,
\eeq
where $c$ is a constant,
$\aaa$ is the lattice spacing, $e$ is the electric charge and the
magnetic charge is \footnote {The notation $g$ is
reserved for the non-Abelian coupling, the magnetic coupling is
denoted as $g_m$.} $g_{m}=1/2e$. Thus, the monopoles are 
infinitely heavy
and, at first sight, this precludes any condensation since the
probability to find a monopole trajectory of the length $L$ is
suppressed as
\beq
\label{action}
\exp(-S)~=~\exp\left(-{c\over e^2}\cdot {L\over \aaa}\right)\,.
\eeq
Note that the constant $c$ depends on the details of the lattice
regularization but can be found explicitly in any particular case.

However, there is an exponentially large enhancement factor due to
the entropy. Namely, trajectory of the length $L$ can be
realized on a cubic lattice in $N_L=7^{L/\aaa}$ various ways. 
Indeed, the monopole occupies center of a cube and the trajectory
consists of $L/a$ steps. At each step the trajectory can be continued
to an adjacent cube. In four dimensions there are 8 such cubes. However,
one of them has to be excluded since the monopole trajectory is
non-backtracking. Thus the entropy factor,
\beq\label{entropy}
N_L~=~\exp\left(\ln 7 \cdot {L\over \aaa}\right)\,,
\eeq
cancels the suppression due to the action
(\ref{action}) if the coupling $e^2$ satisfies the condition
\beq
\label{critical}
e^2_{crit}~=~c/\ln7~\approx~1\,,
\eeq
where we quote the numerical value of $e^2_{crit}$ for the Wilson action
and cubic lattice. At $e^2_{crit}$  any monopole trajectory length
$L$ is allowed and the monopoles condense.

This simple theory works within  about one percent 
accuracy in terms of $e^2_{crit}$
\cite{suzuki1}. Note that the energy-entropy balance above does not
account for interaction with the neighboring monopoles.

\subsection{Monopole cluster in the field-theoretical language}

The derivation of the previous subsection implies that the monopole condensation
occurs when the monopole action is ultraviolet divergent.
On the other hand, the onset of the condensation in
the standard field theoretical language corresponds to 
the zero mass of the magnetically charged field $\phi$.
It is important to emphasize that this apparent
 mismatch between the two languages 
is not specific for the monopoles at all. Actually, there is
a general kinematic relation between
the physical mass of a scalar field $m^2_{phys}$ and the mass
$M$ defined in terms of the (Euclidean) action, $M\equiv S/L$ where
$L$ is the length of the trajectory and $S$ is the corresponding action
\footnote{It is worth emphasizing that the results of the lattice measurements
are commonly expressed in terms of Higgs masses and interaction
constants, see \cite{action}. However, these masses are obtained 
without subtracting the ln7 term (compare Eq (\ref{massrenormalization}))
and, to our belief, are not the physical mass for this reason.
Where by the physical masses we understand the masses in the
continuum limit. In particular, the physical masses determine
the shape of the Abrikosov-like string confining the heavy quarks.}:
\beq\label{massrenormalization}
m^2_{phys}\cdot a~\approx~M-{\ln 7\over a},
\eeq
where terms of higher order in $ma$ are omitted.
Here by $m^2_{phys}$ we understand the mass entering the propagator
of a free particle,
$$D(p^2,m^2_{phys})~\sim~
(p^2+m^2_{phys})^{-1}~,$$
where
$p^2$ is either Euclidean or Minkowskian momentum squared.

In view of the crucial role of the Eq. (\ref{massrenormalization})
for our discussion, let us reiterate the statement.
We consider propagator of a free scalar particle in terms of the
path integral:
\beq\label{pathintegral}
D(x_i,x_f)~\sim~\Sigma_{paths}exp(~-S_{cl}(path)),
\eeq
where for the classical action associated with the path we would like
to substitute simply the action of a point-like classical particle,
$S_{cl}=M\cdot L$ where $M$ is the mass of the particle and $L$ is
the length of the path. Then we learn that there is no such representation 
(with replacement of $S_{cl}$ by $iS_{cl}$))
for the propagator of a relativistic particle in the Minkowski space
because of the backward-in-time motions \footnote{I am indebted to
L. Stodolsky for an illuminating discussions on this topic.}. 
However, in the Euclidean
space the representation (\ref{pathintegral}) works. The physical mass
is, however, gets renormalized compared to $M$ according to (\ref{massrenormalization}). 

Derivation of the Eq (\ref{massrenormalization}) is in textbooks \footnote{
Actually, one  finds mostly $\ln2D\equiv\ln8$ instead of $\ln7$.
We do think that $\ln7$ is the correct number but in fact this difference
is not important for further discussion.} ,
see, e.g.,~\cite{qg}.
The central point is that the action for a point-like particle
in the Euclidean space looks exactly the same as that
of a non-interacting polymer with a non-vanishing chemical potential
for the constituent atoms.
The transition from the polymer to the field theoretical language
is common in the statistical physics (see, e.g., \cite{parisi}).
The first applications to the monopole physics are due to the
authors in Ref.~\cite{stone}. For the sake of completeness 
we reproduce here the main points crucial for our discussion later.
Mostly, we follow the second paper in Ref. \cite{stone}.

The scalar particle trajectory represented as a random walk and the
corresponding partition function is:
\bea\label{randomwalk}
Z = \int \dd^4 \, x \, \sum^\infty_{N=1} \frac{1}{N} \, e^{ - \mu N}
\, Z_N(x,x)\,,
\label{Z}
\eea
where $\mu$ is the chemical
potential and $Z_N(x_0,x_f)$ is the partition function of a polymer
broken into $N$ segments:
\beq
Z_N(x_0,x_f) = \Bigl[\prod\limits^{N-1}_{i=1} \int \dd^4 x_i\Bigr] \,
\prod^{N}_{i=1} \Biggl[\frac{\delta(|x_i - x_{i-1}|-a)}{2\pi^2
a^3}\Biggr]\, \exp\Bigl\{ - \sum\limits^N_{i=1} g
V(x_i)\Bigr\}\,.
\label{ZN}
\eeq
This partition function represents a summation over all atoms of
the polymer weighted by the Boltzmann factors.
The $\delta$--functions in (\ref{ZN}) ensure that each bond in
the polymer has length $a$. The starting point of the polymer
(\ref{ZN}) is $x_0$ and the ending point is $x_f \equiv x_N$.

In the limit $a \to 0$ the partition function (\ref{ZN})
can be treated analogously to a Feynman integral.
The crucial step is the coarse--graining: the $N$--sized polymer
is divided into $m$ units by $n$ atoms ($N = mn$), and the limit is considered
when both $m$ and $n$ are large while $a$ and $\sqrt{n} a$ are
small. We get,
\beq
\label{constraint}
\prod\limits^{(\nu+1)n-1}_{i=\nu n} {1\over
2\pi^2\aaa^3}\delta(|x_i-x_{i+1}|-\aaa) \to
{\Bigl(\frac{2}{\pi n a^2}\Bigr)}^2
\,\exp\Bigl\{ - \frac{2}{n \, a^2} {(x_{(\nu + 1) n} - x_{\nu n
})}^2 \Bigr\} \,,
\eeq
where the index $i$, $i=\nu n \cdots (\nu+1)n-1$, labels the atoms in
$\nu^{\mathrm{th}}$ unit. The polymer partition function
becomes \cite{stone}:
\bea
\label{polymer}
Z_N(x_0,x_f) & = & {\mathrm{const}} \cdot
\Bigl[
\prod^{m-1}_{\nu =1} \dd^4 x\Bigr] \Biggl[
{\Bigl(\frac{2}{\pi n a^2}\Bigr)}^{2 m} \exp\Bigl\{
\sum_{\nu=1}^m{(x_{\nu}-x_{\nu-1})^2\over n\aaa^2}\Bigr\}\Biggr]
\nonumber \\
& & \cdot \exp\Bigl\{ - \sum_{\nu=1}^m n (\mu + V(x_\nu)) \Bigr\}\,.
\eea
The $x_i$'s have been re-labeled so that $x_{\nu}$ is the average
value of $x$ in at the coarser cell. Using the variables:
\beq
\label{related}
s~=~{1\over 8}n\aaa^2\nu,~~~\tau~=~{1\over 8}\aaa^2\,N\,, ~~m_0^2~=~
{8\mu \over \aaa^2}\,,
\eeq
one can rewrite the partition function (\ref{Z}) as
\beq
\label{m0}
Z = {\mathrm{const}}\cdot \int\limits_0^{\infty}
\frac{\dd\tau}{\tau} \, \int\limits_{x(0) = x(\tau)=x}
D x~\exp\Biggl\{-\int\limits_0^\tau \Bigr[{1\over 4}\dot{x}^2_\mu(s)
+ m_0^2 + g_0 V(x(s))
\Bigr] \, \dd s \Biggr\}\,.
\eeq
The next step is to rewrite the integral over trajectories $x(\tau)$ as
the standard path integral representation for a free scalar field.
For us it is important only that the $m_0^2$
term in the Eq. (\ref{m0}) is becoming the standard mass term
in the field theoretical language:
\bea
\cZ & = & \sum\limits_{M=0}^\infty \frac{1}{M!} Z^M \nonumber\\
& = & {\mathrm{const}}\cdot  \int D \phi \, \exp\Bigl\{
- \int \dd^4 x \, \Bigl[ (\partial_\mu \phi)^2 + m^2_0 \, \phi^2
+ g_0 V(x) \phi^2 \Bigr] \Bigr\}\,.
\label{Zphi}
\eea
The whole machinery can be easily generalized to the case of charged
particles (monopoles) with Coulomb-like interactions.

\subsection{Monopole condensation in non-Abelian case: expectations}

If we try to adjust the lessons from the compact $U(1)$
to the non-Abelian case then the good news is that,
indeed, all the $U(1)$ subgroups of the color $SU(2)$
are compact.
Moreover, dynamics of any subgroup of the $SU(2)$
is governed by the same running coupling $g^2(r)$.
Thus, we could hope that the following simple picture
might work:
if the lattice spacing $a$ is small
we would not see monopoles because $g^2(\aaa)$
falls below $e^2_{crit}$. However, going to a
coarser lattice a la Wilson we come to the point
where $g^2(\aaa^2)\approx e^2_{crit}$. Then we
apply the entropy-energy balance which works so well in case
of the compact $U(1)$ and conclude that the monopoles
of a critical size $a_{crit}$ such that $g^2(a_{crit})\sim 1$
condense in the QCD vacuum.

This simple picture is open, however, to painful questions.
First, monopoles are defined topologically within a $U(1)$
subgroup \footnote{Note that a $SU(2)$-invariant definition
of the monopoles is also possible \cite{fedor}. However, their dynamical
characteristics have not beeen measured yet and
such monopoles are not considered here.}. However, it is only the $U(1)$ 
invariant action which has a non-vanishing minimum for
a $U(1)$ topologically non-trivial object. There is no relation,
generally speaking, between the full non-Abelian action and a $U(1)$-subgroup
topology. As an illustration of this general rule, consider \cite{main}
the field configuration generated 
from the vacuum
by the following gauge rotation matrix:
\beq
\label{example}
\Omega~=~\left(\matrix{
e^{i\varphi}\sqrt{A_D} & \sqrt{1-A_D}\cr
-\sqrt{1-A_D}          &  e^{-i\varphi}\sqrt{A_D}\cr
}\right)\,,
\eeq
where $\varphi$ is the angle of rotation around the axis connecting the
monopoles and $A_D$ is the $U(1)$ potential representing pure Abelian
monopole -- antimonopole pair:
\beq
A_{\mu}dx_{\mu}~=~
{1\over 2}\left({z_+\over r_+}-{z_-\over r_-}\right)d\varphi
~\equiv~ A_D(z,\rho)d\varphi\,,
\eeq
where $z_{\pm}=z\pm R/2$, $\rho^2=x^2+y^2$, $r_{\pm}^2=z_{\pm}^2+\rho^2$.
The action associated with the $A_{\mu}^a$ generated in
this way is vanishing since it is a pure gauge. 
In its Abelian part, however, the configuration looks as a
Dirac string with open ends and monopoles at the end points.
It is the ``charged'' vector fields which cancel the contribution
to the non-Abelian field strength tensor
$F_{\mu\nu}^a$ coming from the ``neutral'' field (for details see \cite{main}).

Therefore, there is no reason, at least at first sight,
for the saturation of the functional
integral at the classical solution with infinite action, see
(\ref{linear}).
This observation brings serious doubts on the validity
of our simple dynamical picture.

\section{Monopoles, as they are seen}

\subsection{Monopole dominance}

On the background of the theoretical turmoil, the data
on the monopoles indicate  a very simple and solid picture.
We will constrain ourselves to the monopoles in the so
called Maximal Abelian gauge and the related projection
(MAP). We just mention some facts, a review and further
references can be found, e.g., in Ref.~\cite{review1}.

 Since the monopoles of the non-Abelian theory are expected
to actually be $U(1)$ objects one first uses the gauge freedom
to bring the non-Abelian fields as close to the Abelian
ones as possible. The gauge is defined by maximization of a
functional which in the continuum limit corresponds to $R(\hat{A})$ where
\beq
R(\hat{A})~=~-\int d^4x\big[(A_{\mu}^1)^2+ (A_{\mu}^2)^2\big]
\eeq
where $1,2$ are color indices.

As the next step, one projects the  non-Abelian fields generated on the lattice
into their Abelian part, essentially, by putting $A^{1,2}\equiv 0$.
In this Abelian projection one defines the monopole currents
$k_{\mu}$ for each field configuration. Note that the original configurations
which are used for a search of the monopoles are generated within the
full non-Abelian theory. Upon performing the projection one can
introduce also the corresponding Abelian, or projected action.

The relation of the monopoles to the confinement is revealed
through evaluation of the Wilson loop for the quarks in the
fundamental representation. Namely it turns out, first, that the
string tension in the Abelian projection is close to the string
tension in the original $SU(2)$ theory \cite{mondom}:
\beq
\sigma_{U(1)}~\approx~ \sigma_{SU(2)}\,.
\eeq
Moreover, one can
define also the string tension which arises due to the monopoles
alone. To this end, one calculates the field created by a monopole
current:
\beq
\label{laplasian}
A_{\mu}^{mon}(x) = \frac{1}{2} \varepsilon_{\mu\nu\alpha\beta}
\sum\limits_y \Delta^{-1}(x-y) \, \partial_\nu  m_{\alpha\beta}[y;k]\,,
\eeq
where $\Delta^{-1}$ is the inverse Laplacian, and sums up (numerically)
over the Dirac surface, $m[k]$, spanned on the monopole
currents $k$. The resulting string tension is again close to
that of the un-projected theory:
\beq
\sigma_{mon}~\approx~ \sigma_{SU(2)}\,.
\eeq

It might worth mentioning that these basic features remain also true
upon inclusion of the dynamical fermions in $SU(3)$ case (full lattice
QCD) \cite{latest}.

\subsection{Gauge-invariant properties of the monopoles.}

Despite of the apparent gauge-dependence of the monopoles
introduced within the MAP, they encode gauge-invariant information.
In particular, we would mention two points: scaling of the monopole
density and full non-Abelian action associated with the monopoles.

According to the measurements (see \cite{bornyakov2} and references therein)
the mo\-no\-po\-le density $\rho_{mon}$ in three-dimensional volume (that is, at
any given time) is given in the physical units.
In other words, the density scales according to the
renormgroup as a quantity of dimension 3. Numerically:
\beq
\label{scaling}
\rho_{mon}~=~ 0.65(2) ~(\sigma_{SU(2)})^{3/2}\,.
\eeq
One important remark is in order here. While discussing the
monopole density one should
distinguish between what is sometimes called ultraviolet (UV)  and
infrared (IR) clusters \cite{teper}. The infrared,
or percolating  cluster fills in the whole lattice while the UV clusters
are short. There is a spectrum of the UV clusters, as a function of
their length, while the percolating cluster is in a single copy.
The statement on the scaling (\ref{scaling})
applies only to the IR cluster. We do not consider the UV clusters
in this note.

Also, upon identification of the monopoles in the Abelian
projection, one can measure the non-Abelian action associated with
these monopoles. For practical reasons, the measurements refer to
the plaquettes closest to the center of the cube containing the
monopole. Since the self energy is UV divergent, it might be a
reasonable approximation. The importance of such measurements is
that we expect that it is the non-Abelian action which enters the
energy-entropy balance for the monopoles.

The results of one of the latest measurements of this type
are reproduced in Figure~1 (see \cite{anatomy}).
\begin{figure}
\epsfxsize=7.5cm
\centerline{\epsfbox{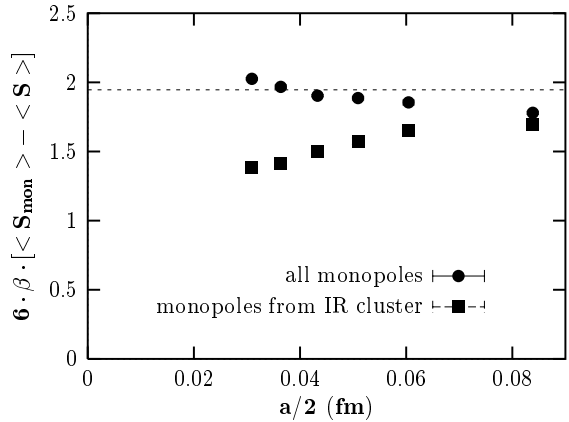}}
\caption{The average excess of the 
full non-Abelian action on
the plaquettes closest to the monopole, as a function of a half of
the lattice spacing $a/2$. The data are reproduced from 
the first paper in Ref. 5}
\label{nonabeact}
\end{figure}
What is plotted here is the average excess of the action on
the plaquettes closest to the monopole (monopoles
are positioned at centers of cubes).
The action is the lattice units. In other words, the
corresponding mass of the
monopole $M_{mon}(a)$ of order $1/\aaa$ if the action of order unit.

As is emphasized in Ref. \cite{anatomy},
the IR and UV monopoles are distinguishable through their
non-Abelian actions. For the UV monopoles the action is larger, in
accordance with the fact that they do not percolate (condense).
This is quite a dramatic confirmation that the condensation
of the monopoles in the Maximal Abelian projection is driven
by the full non-Abelian action, not by its projected counterpart.

\section{Fine tuning}

Let us pause here to reiterate our strategy.
We are assuming that the monopole 
condensation can be described within an effective Higgs-type theory
like (\ref{effective}). 
In fact, even this broad assumption
can be wrong but at this time it is difficult to 
suggest a framework  alternative to the field theory. 
Next, we would like to fix the effective theory
using results of the lattice measurements. Moreover
we are interested first of all in interpreting data 
which can be expressed in gauge independent way.
As the first step, we will argue in this section
that the data on the monopole action \cite{anatomy} imply
a fine tuning. By which we understand that
\beq\label{finetuning}
|M_{mon}(a)-{ln7\over a}|~\ll~M_{mon}(a)
\eeq
where $M_{mon}(a)$ is the monopole self energy \footnote{
We hope that the notations are not confusing: there are two monopole masses
being discussed. One is the standard magnetic field energy
(see (\ref{linear})) and the other is what we call physical mass, $m^2_{phys}$
and this mass determines propagation
of a free monopole.} and $\ln7$ is of pure geometrical origin
(see (\ref{entropy})). Note that (\ref{finetuning})
looks similar to the fine tuning condition in
the Standard Model.

\subsection{Evidence}

There are a few pieces of evidence in favor of the fine tuning (\ref{finetuning}):

a) Direct measurements indicate that the excess of the action is indeed
related to the $\ln7$, as is obvious from Fig. 1. Let us also
emphasize that it is only the full non-Abelian action which ``knows''
about the $\ln7$. The Abelian projected action is not related at all
to the $\ln7$ \cite{anatomy}. This illustrates once again that the
dynamics of the monopoles in MAP is driven by the total $SU(2)$
action.

b) It is difficult to be more quantitative about the excess
of the action basing on the direct data quoted above. In particular,
we should have in mind that for finite $a$ there are 
geometrical corrections to the equation (\ref{entropy}).
Indirect evidence could be more precise. In particular,
it is rather obvious that the scaling of the monopole density
(see Eq. (\ref{scaling})) implies:
\beq\label{assumption}
|M_{mon}(a)- {\ln7\over a}|~\sim ~\Lambda_{QCD}
\eeq
so that the action per unit length of the 
monopole trajectory does not depend on the lattice spacing $a$.

c) Also, independence on the lattice spacing of the temperature (\ref{temperature})
of the phase transition suggests strongly validity of the Eq. (\ref{assumption}).
Indeed, the measurements at the smallest $a$ available, $a\sim 0.06 fm$,
see Fig. 1,  suggest
\beq 
M_{mon}~>~4~GeV, ~~M_{mon}~\gg~T_{deconf},
\eeq
Moreover, it is well known that at the point of the phase transition the
monopole trajectories change drastically.
Such a
sensitivity of the monopoles to the temperature 
is possible only if the effect of the self energy of the monopole is mainly
canceled by the entropy factor, see (\ref{assumption}).

Also, an analysis of the data in Ref. \cite{temperature}
suggests that
\beq
T_{deconf} ~\sim~d_{mon}^{-1},
\eeq
where $d_{mon}$ is the distance between the monopoles in the infrared cluster,
$d_{mon}~\sim~0.5fm$ \cite{anatomy}. Thus the temperature is  not sensitive
 to our ultraviolet parameter which is the size of the monopole.

d) Phenomenological fits suggest \cite{action}:
\beq\label{plus}
\label{mass}
M_{mon}~\approx~M_{mon}^{Coul}(a)+ const, \quad const > 0\,,
\eeq
where
by $M_{mon}$ we understand the action associated with the monopole.
 Note also that the
Coulombic part of the mass, $M_{mon}^{Coul}(a)$ is of order $1/g^2\aaa$.

Let us recall the reader that on the theoretical side our main concern 
was that there is no reason why $M_{mon}(a)$ cannot drop to zero.
Now we see that our fears are not justified: the monopole self energy
is even higher than it would be in the pure Coulomb-like case!
As far as we concentrate on a single monopole there is no way to understand
(\ref{plus}). But this is indeed numerically necessary for the fine tuning.

Thus, the fine tuning (\ref{finetuning}) seems to be granted by the data.

\subsection{The origin of the huge mass scale}

 We are talking actually about small distances, by all the
standards of QCD. The numerical value \cite{anatomy}
 of the size of the monopole (\ref{size})
is much smaller than the inverse temperature of the
phase transition.

The radius (\ref{size}) is defined in terms of
the derivative from the monopole action with respect to $a$,
see \cite{anatomy}. What we would like to emphasize here is that
the actual ``physical size'' of the monopole can be much smaller
than (\ref{size}). By the physical size $R_{phys}$
we  understand now the
distances where the excess of the monopole action is parametrically
smaller than the action associated with the zero-point fluctuations.
It is the $R_{phys}$ where the asymptotic freedom actually reigns,
not $R_{mon}$ quoted in (\ref{size}).

No evidence exists at the moment that reaching $R_{phys}$ is in sight,
see Fig. 1.
Indeed, in the lattice units used in Fig. 1 
the excess of the action density  of order $\Lambda_{QCD}^4$
would look like having zero at $a=0$
and approaching this zero as $a^4$. 
Having in mind the data showed in Fig 1
it is tempting to speculate that the onset of such a behavior is still
far off from the presently available lattice spacings.

Moreover, as we will argue now it looks plausible that
the $R_{phys}$ is shifted to the scale 
\beq\label{rphys}
R_{phys}~\sim ~(100~ GeV)^{-1}~.
\eeq
Before giving arguments in favor of (\ref{rphys}) let us ask
ourselves, why the estimate (\ref{rphys}) is difficult to accept,
at least at first sight so. The reason is obvious: one thinks usually about
non-perturbative effects in quasi-classical terms, which
work in the instanton case. Thus, one assumes that the probability to find
non-perturbative effects is exponentially small at small $g^2(a)$,
$exp(~-c/g^2(a))$.

But the failure of such a logic in the monopole case is
evident from the case of the compact $U(1)$, see above.
Even the monopoles with infinite (Euclidean) action condense.
Moreover, $R_{phys}$ is naturally determined by the running
of the coupling which is logarithmic and can result
in huge factors in the linear scale.

Let us make simple estimates. Namely, the $U(1)$
critical coupling is well known, $e^2_{crit}\sim 1$. In the QCD case we can
rewrite the condition (\ref{critical}) as a condition on the $R_{phys}$.
In the realistic case
we have at the LEP energies $E^2 \sim (100~\mbox{GeV})^2$, $\alpha \approx
0.1$. Then
\beq\label{huge}
M_{phys}~\sim~\mbox{TeV}
\eeq
and, remarkably enough,
we are getting rather the weak interactions scale than $\sim\Lambda_{QCD}$.

Also, the $SU(2)$ lattice measurements typically refer to $\beta\sim 2.6$ while
our guess about $R_{phys}$ asks for measurements at $\beta\sim 4$
which are absolutely unrealistic at the moment.

Thus, we come to a paradoxical conclusion that
the presently available $\beta$ are too low to see
dissolution of the monopoles at small distances. 
Moreover, because the running of the coupling
is only logarithmic the scale of of the onset
asymptotic freedom -- which is defined now as the vanishing of
the excess of the monopole action compared to the zero-point-fluctuations
action-- can be very far off.    

It is amusing to notice \footnote{The observation is due to M.I. Polikarpov.}
that in case of the $SU(3)$ gluodynamics on the lattice
$g^2=1$, or $\beta=6$ corresponds to the lattice
spacing $a\approx 0.1\,\,\,fm$ and the scale is:$$
R^{SU(3)}_{phys}~\sim~(2\mbox{GeV})^{-1}\,.$$
Thus, through dedicated studies of the monopoles in the $SU(3)$ case it is possible
to clarify whether there is a crucial change
in the monopole structure at the point $g^2(a)\approx 1$. 

\subsection{Supersymmetry}
 
We are pursuing a pure phenomenological approach 
and are not in position now to discuss possible
mechanisms ensuring the mass hierarchy within the
effective scalar filed theory. Obviously, it is
not a simple question. The same obvious, the supersymmetry could be
an answer
\cite{seiberg,espriu}. 

Generically, the supersymmetry would imply
that there are magnetically charged fields with spin $1/2$ as well.
Spin of the magnetically charged 
particles can be determined from the character of their trajectories.
The random-walk representation (\ref{randomwalk}) is true only
for the scalar particles. For spinors, there is an intrinsic rigidity
\cite{qg}. To detect the rigidity, one can measure the correlation function
between the vectors tangent to the trajectory.
 
Note that we expect that the particles in the IR cluster are 
scalars for sure. On the other hand, UV , or relatively short trajectories
could correspond both to scalar and spinor particles. Detecting spinor particle 
propagating on the lattice would be a spectacular indication to the
supersymmetry. And vice versa.

\section{Naive limit $\aaa\to 0$}

If we get convinced that there might exist mass hierarchy
then we come to the next question which seems even more difficult.
Namely, why there is no independent phenomenological evidence for
the existence of large ``ultraviolet'' mass scale, like (\ref{rphys}).
Indeed, only at this scale we are guaranteed that zero-point fluctuations
dominate over the non-perturbative (monopole) fluctuations.
In an attempt to answer this question let us consider limit $a\to 0$
assuming that in this limit we are still having the same behavior
of the monopole action as at the presently available lattices.
If, indeed, $R_{phys}$ is as small as indicated by, say, (\ref{rphys})
then the validity of the approximation $a\to 0$ seems granted.

\subsection{Power-like dependences on the lattice spacing $a$}
 
Coming back to the partition function (\ref{m0}),
the monopole condensation corresponds to a negative $m^2_0$.
The physical excitations should be redefined in terms
of the new vacuum. The standard strategy to study these excitations
is to measure various vacuum correlators of the field $\phi$.
At present time, however, there is a lot of data on the vacuum
fields, also in terms of the monopole trajectories,
but not on the correlators. 
There is no rigorous way to interpret these data.
Still, at least naively, one can relate some of the vacuum characteristics
to derivatives from the partition function with respect
to the parameters, such as $\mu$, $m_0^2$. The idea goes back to
the first paper in Ref. \cite{stone}. We supplement this idea by
the knowledge of the properties of the infrared monopole cluster,
which represents the non-perturbative vacuum in our
picture.

To get a relation for $\rho_{mon}$ 
let us differentiate first
the partition function in the polymer
representation with respect to the chemical potential $\mu$:
\beq
\label{l}
\langle L \rangle = \frac{\partial}{\partial \mu} \ln \cZ\,
\eeq
where $L$ is the length of the monopole trajectory.
Since  the density $\rho_{mon}$ scales:
\beq
\langle L \rangle = \rho_{mon}\cdot V_4\,,
\eeq
where $V_4$ is the 4-volume occupied by the lattice.
On the other hand, differentiating the same partition function but
in the field theoretical representation (\ref{m0}) with respect to
$m_0^2$ we get the vacuum condensate:
\beq\label{phi}
\langle \phi^2\rangle = \frac{\partial}{\partial m^2_0} \ln \cZ\,.
\eeq
It is worth emphasizing that in the both cases (\ref{l}) and (\ref{phi})
we keep only the contribution of the IR monopole cluster corresponding
to the condensing Higgs field in the field-theoretic language.

Finally, since the parameters $\mu$ and $m_0^2$ are directly related, see
Eq. (\ref{related}), we get:
\beq\label{mainn}
\langle \phi^2\rangle~=~{1\over 8}\rho_{mon}\cdot \aaa~,
\eeq
which is one of our main results.
Note that,
up to an overall numerical factor,
Eq. (\ref{mainn}) is quite obvious on the dimensional grounds.

Thus, let us assume that the scaling of the monopole density 
in the IR cluster $\rho_{mon}$
continues to be true for smaller lattice spacings as well,
at least until we reach the mass scale sensitive to
the non-local structure of the monopoles,
see discussion above. Then we have the following simple picture:
\beq\label{a.f.}
\lim\limits_{\aaa\to 0}{m_0^2}~\sim~{\mu\over \aaa}~\rar~\infty,~~
\lim\limits_{\aaa\to 0}{\langle \phi^2\rangle}~\sim~\rho_{mon}\aaa~\rar~0,~~
\lim\limits_{\aaa\to 0}{m_V^2}~\sim~g^{-2}\rho_{mon}\aaa~\rar~0.
\eeq
It is worth emphasizing that the masses we are discussing here are
gauge invariant since we started from the
non-Abelian action per unit length. And we see
that existence of the huge mass scale (\ref{huge}) might in fact be in
no contradiction with the asymptotic freedom. Indeed, only the
chemical potential has physical meaning
and the scaling of the $\rho_{mon}$ indicates that
it is of order $\Lambda_{QCD}$. Moreover,
the effect of the condensate on the gluon mass goes away as a power of $\aaa$.

It is worth emphasizing that Eq. (\ref{a.f.}) implies that
\beq
\lim\limits_{\aaa\to 0}{m_0^2\cdot \langle \phi^2\rangle}~\sim~\mbox{const}\,.
\eeq
In other words, the potential energy behaves smoothly as $\aaa\to 0$.
And this is, in fact, the most adequate formulation of the
emerging picture. It was possible to find the $\aaa$-dependence for $m_0^2$
and  $\langle \phi^2\rangle$ separately only because of normalizing
the kinetic energy to unit.

Note that the scaling laws (\ref{a.f.}) are still consistent with
$\rho_{mon}=~const$. Moreover, this seems to be sufficient
to ensure the monopole dominance and
\beq
\lim\limits_{\aaa\to 0}{\sigma_{mon}}~\sim ~\mbox{const}\,,
\eeq
where the monopole string tension is calculated with the use
of Eq. (\ref{laplasian}). Which means in turn that the parameters
used to describe the structure of the string within the Abelian
projection can be stable in the limit $\aaa~\to ~0$.
Moreover, say,
\beq
\lim_{\aaa\to 0} {(m_V^2)_{Ab. proj.}}~\sim~ \mbox{const}\,,
\eeq
is in no  direct contradiction with (\ref{a.f.}) since the masses
determined in terms of the Abelian-projected action are not directly
related to the masses (\ref{a.f.}) determined
in terms of the non-Abelian action. 

Thus, the picture which emerges if we start with assumption (\ref{assumption})
has some attractive features. In particular, it removes the ultraviolet
scale from observables in an amusingly simple way. However, our estimates
are indeed naive and the discussion is preliminary.

\subsection{Phenomenology}

Studying characteristics of the monopole trajectories on
the lattice provides with a unique possibility
to visualize field theory in the polymer representation.
We have already seen that the measurements of the
$SU(2)$ invariant action allowed for far reaching conclusions
on the underlying Higgs-type models.

Let us list some predictions which could be checked directly on the lattice:

a) The monopole trajectories are random walk for any $\aaa$
in the sense that there is no correlation between the vectors
tangent to the monopole trajectory. This is true for scalar particles.
As we mentioned above, it is important to check this prediction 
both for IR and UV monopole clusters. 

b) monopole density scales, $\rho_{mon}=const$ and is independent
of $a$ at least as far as the monopole action exceeds the average
in the lattice units (and not in $\Lambda_{QCD}^4$).

c) as is known (see, e.g., \cite{intersect})
the monopole trajectories intersect.
It is natural to speculate that the distance between the self-intersections
also scales, reflecting the scaling of the potential energy.

d) The intersections correspond in the field theoretical language to the
$\lambda\phi^4$ interaction:
\beq
V(\phi)~=~-m_0^2\phi^2 ~+~\lambda \phi^4
\eeq
As we argued, one expects that the potential
energy is $\aaa$-independent. This would imply that the effective scalar mass
defined in terms of the second derivative of the potential at
the minimum is also $\aaa$-independent. Which could be checked through
measurements.

e) It would be most interesting to try 
to generate the monopole trajectories within the polymer approach and compare the
results with the simulations within the full QCD. 
In the simplest version, 
there are essentially two entries in the action in the polymer approach,
that is the  chemical potential and the interaction which
is presumably Coulomb-like:
\beq\label{nou1}
S~=~L\mu + g^2_{m}\, {\sum_{a,b}}' {\aaa^2\over(r_a-r_b)^2}\,,
\eeq
where the primed sum,  $\Sigma^{'}_{a,b}$, does not include the self-energy.

\section{Conclusions}

We have argued that data are emerging which indicate that QCD, when
projected onto the scalar-field theory via monopoles corresponds to a
fine tuned theory. Which is if course extremely interesting, if true,
in view of the mystery of the fine tuning in the Standard Model.
The monopoles which we considered are defined (``detected'')
through the Maximal Abelian projection. However, the mass scales which
exhibit mass hierarchy are gauge independent. The scales are provided by
the $SU(2)$ invariant action per unit length of the monopole trajectory,
on one hand, and by the temperature of the phase transition, on the other.  
More generally, we have found that the polymer approach allows to get a
new insight into the mechanism of the monopole condensation. 

\section*{Acknowledgements}

I am grateful to S. Caracciolo, 
M.N. Chernodub, F.V. Gubarev, R. Hofmann, K. Konishi, K. Langfeld, S. Narison,
M.I. Polikarpov and L. Stodolsky  
for discussions. Special thanks are due to M.N. Chernodub for numerous
communications and thorough discussions of the results.

The paper was worked out mostly during the author's stay at
the Scuola Normale Superiore in Pisa.
The hospitality extended to the author by
the members of the Theory group and especially by Prof. R. Barbieri
is gratefully acknowledged.

\section*{References}
\addcontentsline{toc}{section}{\numberline{}References}

\end{document}